\documentstyle[12pt]{article}

minus 1pt

\def\Journal#1#2#3#4{{#1} {\bf #2}, #3 (#4)}

\def\NCL{\em Nuovo Cimento Letters}

\def\NPB{{\em Nucl. Phys.} B}
\def\PLB{{\em Phys. Lett.}  B}
\def\PRL{\em Phys. Rev. Lett.}

\def\JETPL{\em JETP Lett.}
\def\EPJC{{\em Eur.Phys.J.} C}

\def\be{\begin{equation}}\def\ee{\end{equation}}
\def\bea{\begin{eqnarray}}
\def\eea{\end{eqnarray}}

\title{\bf THE ODDERON IN THEORY AND EXPERIMENT\\A MINI-REVIEW}

\author{Basarab NICOLESCU\\LPTPE, Universit\'e Pierre et Marie Curie\\
4,Place Jussieu, 75252 Paris Cedex 05, France\\
E-mail: nicolesc@ipno.in2p3.fr\\\\\\
Talk at the XXIX International Conference on High Energy Physics,\\
Vancouver,Canada, July 23-30, 1998\\ (to be published in the Proceedings of
this Conference)}
\date{}
\def\beq{\begin{equation}}
\def\eeq
{\end{equation}}
\def\noi{\noindent}

\begin{document}
\maketitle
\medskip

\noi{\bf
Abstract.}
We review recent theoretical and phenomenological results on both the
perturbative and non-perturbative Odderon. The HERA type of experiments
constitutes a direct probe of the Odderon.
\vspace {4.5cm}

\noi{\Large\bf LPTPE/UP6/98-13
\hspace{5cm}October 1998}
\newpage


\section{Introduction}
The concept of Odderon was introduced in 1973 \cite{lu}. The Odderon is a
J-plane singularity near J=1 in the odd-under-crossing amplitude $F_-$. It
was first formulated in the framework of asymptotic theorems, i.e. rigorous
results derived from general principles - analyticity, unitarity and
positivity. Its invention was stimulated by the experimental discovery, at
ISR in 1972, of increasing total cross-sections.

The Odderon was longtime considered as an heretical concept because of the
belief that $F_-$ is dominated , at high energies , by $J=1/2$ singularities
($\rho$ and $\omega$ Regge poles). However, it was rediscovered in 1980 in QCD
, where it appears as a compound state of 3 reggeized gluons
\cite{ba}. From
1990, there were important theoretical developments in perturbative QCD. There 
are also detailed phenomenological studies of the non-perturbative Odderon
\cite{ga}. Finally, there is a running experiment ar HERA, constituting a
direct probe of the Odderon \cite{h1}.

Of course, the Odderon has a long history. I will discuss here only recent
results, some of them presented at the Heidelberg workshop ``Pomeron and
Odderon in Theory and Experiment'' \cite{http}.


\section{Calculation of the intercept of the perturbative Odderon in QCD}
There are two lines of research in this framework:

1.The equivalence, in the multi-colour limit, with the hamiltonian of the
completely integrable one-dimensional Heisenberg magnet \cite{li}.

2.Variational methods combined with conformal invariant techniques.

In the first line of research, the basic idea is that the Odderon is a C-odd
state of 3 reggeized gluons which interact pairwise with a well-defined
potential.The problem is to find an operator  $\hat{q}_3$,
\be
\hat{q}_3^2 = - r_{12}^2 r_{23}^2 r_{31}^2 p_1^2 p_2^2 p_3^2  , 
\label{eq:br1}
\ee
which commutes with the Odderon hamiltonian H,
\be
[ \hat{q}_3^2,H ] = 0
\label{eq:br2}
\ee
and has a much simpler form than H.

Recently, Janik and Wosiek \cite{ja} did find the exact solution of the
problem, formulated in terms of the eigenequation
\be
\hat{q}_3 f =q_3 f,
\label{eq:br3}
\ee
where
\bea
f &=& \left(\frac {\rho_{12} \rho_{13} \rho_{23}} 
{\rho^2_{10} \rho^2_{20} \rho^2_{30}}\right)^\mu \Phi(z),   
\label{eq:br4}
\eea
with $\mu=h/3$, h being the conformal weight.
The function $\Phi$  
satisfies a third order linear equation
\bea
a(z) {d^3 \over d z^3}\Phi(z) + b(z){d^2\over d z^2} \Phi(z) 
+c(z){d\over dz}\Phi(z) \nonumber \\ \noalign{\vskip 3pt} + d(z) \Phi(z)=0,
\label{eq:br5}
\eea
where
\be
a(z)=z^3 (1-z)^3,
\label{eq:br6}
\ee

\be
b(z)=2 z^2 (1-z)^2 (1-2z),
\label{eq:br7}
\ee

\be
c(z) = z(z-1)\left(z(z-1)(3\mu+2)(\mu-1)+3\mu^2-\mu\right),
\label{eq:br8}
\ee

\be
d(z)=\mu^2 (1-\mu)(z+1)(z-2)(2z-1) -iq_3 z(1-z),
\label{eq:br9}
\ee 

\be
z=\frac {\rho_{12} \rho_{30}} {\rho_{10} \rho_{32}},
\rho_k=x_k+iy_k (k=1,2,3), \rho_{ij}=\rho_i - \rho_j.   
\label{eq:jw3}
\ee

Janik and Wosiek did find the solution
\be
q_3=-0.20526\,i,
\label{eq:br10}
\ee
corresponding to the Odderon energy
\be
\epsilon=0.16478.
\label{eq:br11}
\ee

The relation between the Odderon energy $\epsilon$
and the Odderon intercept $\alpha_O(0)$
is given by the equation
\be
\alpha_O(0) = 1-(9\alpha_s/2\pi)\epsilon.
\label{eq:br12}
\ee

For realistic values of $\alpha_s$  ($\alpha_s \simeq 0.19$)
one gets

\be
\alpha_O(0) = 0.94.
\label{eq:br13}
\ee

The second line of research was formulated in Ref.8. Recent results were
obtained in Ref.9, where a direct calculation of the lower bound for the
Odderon intercept was performed.

The Odderon energy is defined as
\be
\epsilon = E/D .
\label{eq:br14}
\ee

In Eq.(15) D is a normalization constant and the energy functional is given
by
\be
E=\sum_{n=-\infty}^{\infty}\int_{-\infty}^{\infty}d\nu\epsilon_{n}(\nu)
|\alpha_{n}(\nu)|^{2},
\label{eq:br15}
\ee 
where
\be
\epsilon_{n}(\nu)=2 Re[\psi(\frac{1+|n|}{2}+i\nu)-\psi(1)],
\label{eq:br16}
\ee 

\bea
\alpha_{n}(\nu) &=& \int_{0}^{\infty} dr r^{-2-2i\nu}\int_{0}^{2\pi} d\phi  
e^{-in\phi} \nonumber \\ \noalign{\vskip 3pt} &&{}
\left(i\nu+\frac{n+1}{2}+re^{i\phi}(h-i\nu-\frac{n-1}{2})\right)
(i\nu-\frac{n-1}{2}) \nonumber \\ \noalign{\vskip 3pt} &&{}
(-\tilde{h}+i\nu-\frac{n-1}{2}) Z(r,\phi),
\label{eq:br17}
\eea

\bea
h &=& 1/2+n/2-i\nu, \tilde{h} = 1/2-n/2+i\nu, \nonumber \\ 
 \noalign{\vskip 3pt} &&{} - \infty<\nu<\infty, n=\ldots ,-1,0,1, \ldots ,
\label{eq:br18}
\eea

\be
Z(z,z^*)=|z(1-z)|^{2h/3}\Psi(z,z^*).
\label{eq:br19}
\ee

The function $\Psi$
in Eq.(20) is invariant under the transformations $z\rightarrow 1-z$ and
$z\rightarrow 1/z$
(Bose symmetry in the 3 gluons).

The following trial function $\Psi$
was used in Ref.9:

\be
\Psi = \sum_{k=1}^{N-N_1} c_k a^{k/2-1/6}
+\sum_{k=1}^{N_1} d_k a^{k-1/6}\ln a,
\label{eq:br20}
\ee

where

\be
a=\frac{r^2r_1^2}{(1+r^2)(1+r_1^2)(r^2+r_1^2)} .
\label{eq:br21}
\ee

The result is

\be
\epsilon=0.22269,
\label{eq:br22}
\ee
a value which has to be compared with Eq.(12).
The corresponding Odderon intercept is

\be
\alpha_O(0)= 0.96.
\label{eq:br23}
\ee

By comparing the values (23) and (14), one sees that there is only a 2\%
difference between the ``exact'' result and the variational one.

We draw from this section the conclusion that the Odderon intercept is very
close to 1, i.e. is much higher than the 1/2 ($\rho$,$\omega$) 
intercept. We therefore expect important Odderon effects at high energy.

\section{Recent phenomenological studies of the non-\\perturbative Odderon}
Kilian and Nachtmann \cite{ki} recently shown that the Odderon could induce
spectacular effects in the pseudoscalar meson production in ep scattering at
high energies,e.g. in the $p_\perp$
and rapidity distributions for pion production in the photoproduction
region. The model used in Ref.10 - a simple Regge pole contribution - is
nice as a toy model but is highly unrealistic : a detailed study of $\rho$
values ($\rho = ReF/ ImF$)
at low and medium energies \cite{ga} shows beyond doubt that the Odderon
Regge pole has only a very small contribution, at least in the forward
direction. The forward and non-forward data in a huge energy range favor the
maximal Odderon \cite{ga}. On a strictly theoretical level, the Regge
singularity in the perturbative QCD is much more complicated than a simple
Regge pole and it very difficult to imagine the dynamical mechanism through
which such a singularity could be resolved in just a simple Regge pole in
the non-perturbative region.

Another recent phenomenological study of the Odderon originates from the
Stochastic Vacuum Model (SVM) of Dosch and Simonov \cite{do1,do2}.
A very interesting connection can be established in SVM between the Y-shape
of the baryon and the coupling of the Odderon. The particular diquark-quark
structure of the baryon corresponds to the suppression of the Odderon as a
simple Regge pole which, as we already discussed, is not favored by the data. 
Fortunately, the authors of Ref.12 are able to make one prediction which is
quasi-independent on the singularity structure of the Odderon: a high
cross-section (of the order of 300 nb) in the photoproduction of pions with
single dissociation (breakup of the target proton). The HERA type of
experiments could check this interesting prediction.

\section{Odderon and experimental data}
A natural difficulty in detecting Odderon effects is the fact that in
general the Odderon ($F_-$ amplitude) is mixed with the Pomeron ($F_+$
amplitude). Moreover, the experimental fact that the difference of
antihadron-hadron and hadron-hadron total cross-sections
$\Delta \sigma \propto ImF_-$ is much smaller than the total
cross-section itself, $\sigma_{T} \propto ImF_+$,
indicates that the coupling of the Odderon is much smaller than the coupling
of the Pomeron.It is therefore not surprising that, till now, there are only
two relatively clear experimental indications of the Odderon:

1) the experimental discovery at ISR \cite{isr} of a difference between
$(d\sigma/dt)_{\footnotesize \bar{p}p}$ and
$(d\sigma/dt)_{\footnotesize pp}$
in the dip-shoulder region ($|t| \simeq 1.3\:GeV^2$)
.This difference is induced by an Odderon with an effective intercept close
to 1, a fact which nicely fits the prediction \cite{ga1,br} of the 
perturbative QCD. Moreover,the extrapolation of the maximal Odderon at high t
nicely reproduces the existing data \cite{ga};

2) the experimental discovery of a bump at very small t
($|t| \simeq 2\cdot 10^{-3}\:GeV^2$)
in the high precision dN/dt $\bar{p}{p}$ data \cite{ua} at $\sqrt{s}=541\: GeV$
can be interpreted in terms of oscillations of a very small
period, which could be related to unitarity constraints. In their turn,
these oscillations induce a high value of the semi-theoretical parameter $\rho$
at $t=0$, compatible with an important Odderon contribution in the forward
direction \cite{se}.

However,the most convincing evidence of the Odderon could come from the
pseudoscalar meson production at HERA. The H1 experiment \cite{h1}
offers an unique opportunity of exploring pure $C=-$  channels, without any
Pomeron mixing: the Odderon is here in competition only with the photon. The
results of the experiment will be available by the beginning of 1999.

\section{Conclusions}
There is now quite an intense activity concerning both the perturbative and
the non-perturbative Odderon. QCD calculations show that the intercept of the
perturbative Odderon is very close to 1. The non-perturbative Odderon has
also to be visible in the HERA type of experiments.

\section*{Acknowledgements}
I thank Dr. Pierre Gauron for very useful discussions.

\section*{}


\begin{thebibliography}{99}
\bibitem{lu}L.Lukaszuk and B.Nicolescu, \Journal{\NCL} {8} {405} {1973}.
\bibitem{ba}J.Bartels, \Journal {\NPB} {175} {365} {1980}; J.Kwiecinski and
M.Praszalowicz, \Journal {\PLB} {94} {413} {1980}.
\bibitem{ga} P.Gauron, E.Leader and B.Nicolescu, \Journal{\PLB} {238} {406}
{1990}.
\bibitem{h1} H1 Collab., S.Tapprogge et al. in Proc. of the Int. Conf. on the
Structure and the Interactions of the Photon (Photon 92), ed.F.C.Erne and
A.Bujis (World Scientific, Singapore, 1993).
\bibitem{http} http://www.thphys.uni-heidelberg.de/ws/
\bibitem{li} L.N.Lipatov, \Journal{\JETPL} {59} {571} {1994}; L.D.Fadeev and
G.P.Korchemsky, \Journal{\PLB} {342} {311} {1994}; G.P.Korchemsky,
\Journal{\NPB} {443} {255} {1995}; \Journal{\NPB} {462} {333} {1996}.
\bibitem{ja} R.A.Janik and J.Wosiek, hep-th/9802100; see also L.N.Lipatov,
talk given at IPN Orsay, France, March 1997.
\bibitem{ga1} P.Gauron, L.N.Lipatov and B.Nicolescu, \Journal {\PLB} {260}
{407} {1991}; \Journal {\PLB} {304} {334} {1993}.
\bibitem{br} M.A.Braun, P.Gauron and B.Nicolescu, hep-ph/9809567.
\bibitem{ki} W.Kilian and O.Nachtmann, \Journal {\EPJC} {5} {317} {1998}.
\bibitem{do1} H.G.Dosch and Y.A.Simonov, \Journal {\PLB} {205} {339} {1988}.
\bibitem{do2} M.Rueter and H.G.Dosch, \Journal {\PLB} {380} {177} {1996};
M.Rueter, H.G.Dosch and O.Nachtmann, hep-ph/9806342.
\bibitem{isr} A.Breakstone {\it et al.}, \Journal {\PRL} {54} {2180} {1985};
S.Erhan {\it et al.}, \Journal {\PLB} {152} {131} {1985}.
\bibitem{ua} UA4/2 Coll., C.Augier {\it et al.}, \Journal {\PLB}
{316} {448} {1993}.
\bibitem{se} P.Gauron, B.Nicolescu and O.V.Selyugin, \Journal {\PLB} {397}
{305} {1997}.
\end{thebibliography}
\end{document}